\documentclass[twocolumn,aps, reprint, showpacs,preprintnumbers,amsmath,amssymb,superscriptaddress]{revtex4-1}

\usepackage[pdftex]{graphicx}
\usepackage{amsmath}
\usepackage{fancyhdr}
\usepackage{amsfonts}
\usepackage{amssymb}
\usepackage{color}
\usepackage{float}
\usepackage{color}
\usepackage{natbib}
\begin{document}
\title{Role of dual nuclear baths on spin blockade leakage current bistabilities}
\bigskip
\author{Siddharth Buddhiraju}
\author{Bhaskaran Muralidharan}
\affiliation{Center of Excellence in Nanoelectronics, Department of Electrical Engineering, Indian Institute of Technology Bombay, Powai, Mumbai-400076, India}
\date{\today}
\medskip
\widetext
\begin{abstract}
Spin-blockaded electronic transport across a double quantum dot (DQD) system represents an important advancement in the area of spin-based quantum information. The basic mechanism underlying the blockade is the formation of a blocking triplet state. The bistability of the leakage current as a function of the applied magnetic field in this regime is believed to arise from the effect of nuclear Overhauser fields on spin-flip transitions between the blocking triplet and the conducting singlet states. The objective of this paper is to present the nuances of considering a two bath model on the experimentally observed current bistability by employing a self consistent simulation of the nuclear spin dynamics coupled with the electronic transport of the DQD set up. In doing so, we first discuss the important subtleties involved in the microscopic derivation of the hyperfine mediated spin flip rates. We then give insights as to how the differences between the two nuclear baths and the resulting difference Overhauser field affect the two-electron states of the DQD, and their connection with the experimentally observed current hysteresis curve.
\newline\newline
This is an author-created, un-copyedited version of the article accepted for publication in the Journal of Physics: Condensed Matter. Please see page footers for details. 

\end{abstract}
\maketitle
\section{Introduction}

\begin{figure}
	\centering
		\includegraphics[width=2.7in,height=2.4in]{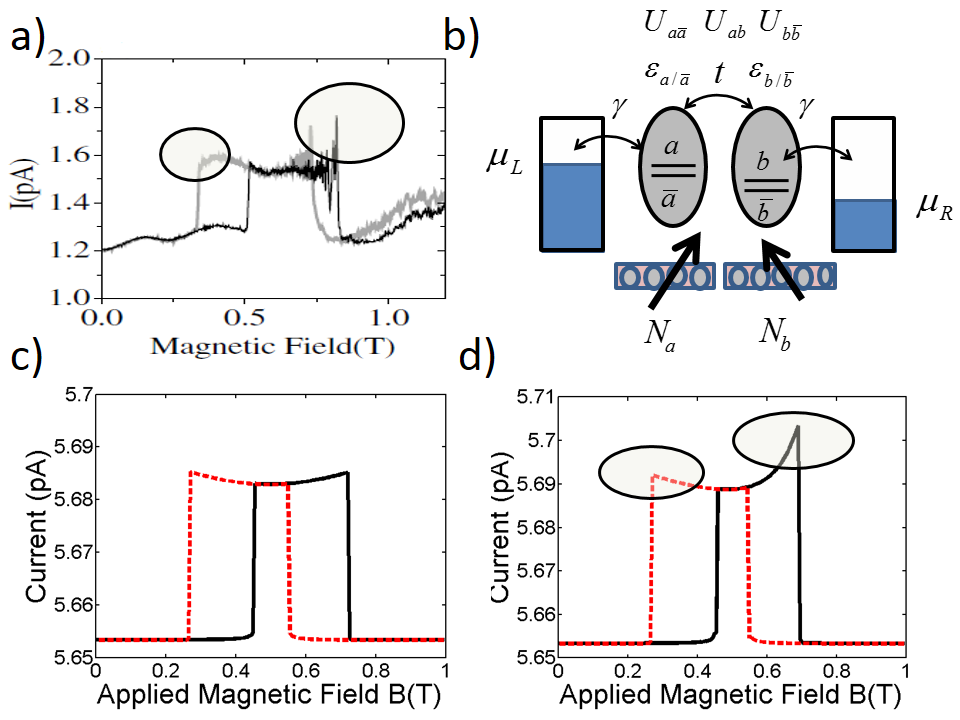}
		\caption{ a) Experimental trace of the $I-B$ characteristic obtained in \cite{Ono} b) Schematic of the DQD structure whose individual dots are labeled $A$ and $B$. Each dot is coupled to its own host nuclei which may in general have $N_a$ and $N_b$ nuclei respectively. The DQD Hamiltonian comprises the on-site energies, hopping energies and the Coulomb interaction matrix. Each nuclear bath may be represented by collective nuclear polarization variables $F_I^a$ and $F_I^b$ respectively. c) Simulated $I-B$ characteristic without a difference Overhauser field d) with the difference Overhauser field. Notice the fin structure developed at the terminal edges of the characteristic that is asymmetric in height similar to the experimental trace in a).}
\label{Fig1}
\end{figure}

Spin-blockaded electronic transport across a GaAs double quantum dot (DQD) system \cite{Tarucha_1} represents an important advancement in the area of spin-based quantum information \cite{Leo_review}. The origin of this blockade is attributed to the blocking triplet state in the two-electron spectrum since it can be filled but not emptied easily \cite{Basky_Datta}. Hyperfine interaction with the host nuclei mediates the lifting of this blockade via triplet-singlet spin-flips resulting in leakage currents. Feedback with host nuclei results in non-trivial leakage current bistabilities \cite{Ono} and many other complex temporal phenomena \cite{SB_1,SB_2,SB_3,Yakoby}. As a result, there has been a surge of theoretical research in the area of hyperfine mediated electronic transport typically predicting novel feedback related phenomena \cite{Levitov_1,Levitov_2,Levitov_3}, exploring electronic state control \cite{Nazarov}, studying relaxation effects \cite{Danon,Inarrea,Taylor}, dephasing mechanisms \cite{Taylor}, and regimes of operation \cite{Lukin}. The focus of this paper, however, is to convey a deeper insight of the hysteresis of the leakage current with respect to an applied magnetic field \cite{Ono} as shown in Fig.~\ref{Fig1}(a).
\\ \indent  A realistic picture of the DQD set up has the electronic states delocalized/entangled over the two dots while the nuclear spins are localized on each dot as shown in the schematic of Fig.~\ref{Fig1}(b). This can only be correctly captured by coupling DQD electronic transport with the dynamics of the two isolated nuclear baths using two separate nuclear polarization variables. In previous studies \cite{Platero_1,Platero_2}, the use of only one nuclear polarization variable made it mandatory for triplet-singlet spin flips to be activated only in the presence of different hyperfine parameters. The idea of using two nuclear baths has also been considered in past works \cite{Levitov_3, Danon_PRL}, however, with specific focus on interpretation of certain non-trivial temporal dynamics, and in \cite{Inarrea}, with specific focus on the DQD energetics and related I-V curves. 
\\ \indent The principal contribution of this paper is to elucidate the non-trivialities that arise when two nuclear bath variables and their dynamics are employed in conjunction with the detailed electronic structure of the DQD system, with particular emphasis on the interpretation of the bistability noted in \cite{Ono}. This is done via a self consistent simulation of nuclear spin dynamics on the two separate baths coupled with a detailed model for the electron transport across the DQD system. With this setup and using feedback-coupled electronic resonance models, we show that even without the inclusion of statistical fluctuations in the nuclear field, equal Overhauser fields on the two dots give rise to bistability as shown in Fig.~\ref{Fig1}(c), while the inevitable differences between dots that result in a difference Overhauser field, lead to additional features such as the observed {\it{fin structure}} at the edges of the hysteresis curve, shown in Fig.~\ref{Fig1}(d). We then briefly discuss the role of the DQD electronic structure which may, in addition, contribute to similar deviations from the flat-topped nature of the $I-B$ trace. Apart from the hysteresis, \cite{Ono} also features an unstable region marked by oscillatory behaviour at higher fields, which is not considered within this model.
\\ \indent In what follows we first focus on the generic formulation that would be used to describe the set up. Following this, we develop the understanding of the origin of the hysteresis using a simple toy model, that describes its essence based on the phenomenon of resonance dragging, after which we will describe the flat topped structure as a result of a dual resonance among the electronic Fock states \cite{Levitov_1} of the double quantum dot. The role of the difference Overhauser field in terms of the experimental signature of the fin structure is set to be discussed in the final subsection before concluding. \\
\section{General Formulation}
\begin{figure}
	\centering
		\includegraphics[width=3.6in,height=3.8in]{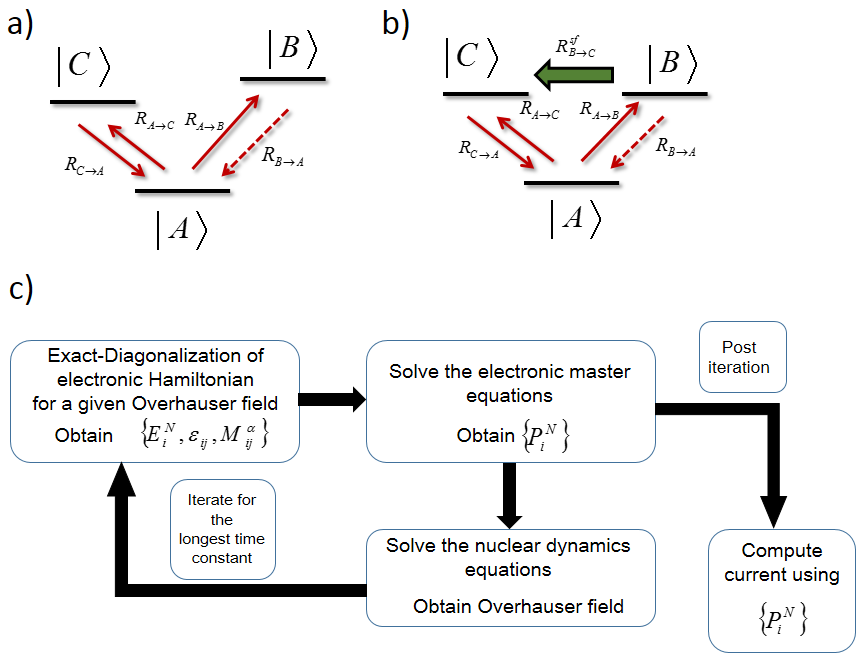}
		\caption{a) Generic three state model for spin blockade with $|B\rangle$ being the blocking state, $|C\rangle$ being the conducting state and $|A\rangle$ being the ground state. The rates $R_{A\rightarrow C}$ etc., are transition rates. b) The same setup with $|C\rangle$ and $|B\rangle$ in resonance as a result of an applied magnetic field. Spin-flip process from the blocking to the conducting state are now activated and are represented with a green arrow. c) Flow chart describing the self-consistent procedure for solving such a problem at a given external magnetic field. Usually the longest time constant is the nuclear-spin relaxation time.}
\label{Flow}
\end{figure}

In this section, we elucidate the basic formulation using which we can describe the hyperfine mediated transport across a spin blockaded set up.  We first consider a general situation of spin blockade \cite{Basky_Datta}, which can be phenomenologically described by the three state model considered in Fig. \ref{Flow}(a). We consider a bias regime in which transport is described by tunneling transitions between the multi-electron state $|A\rangle$  and the states $|B\rangle$ and $|C\rangle$. The state $|B\rangle$ is a blocking state whereas $|C\rangle$ is conducting, both of which belong to an $N$ electron subspace, and the state $|A\rangle$ is in an $N-1$ electron subspace. The arrows indicate transitions between the states. The blocking state is characterized by a slow rate of transition between $|B\rangle$ and $|A\rangle$ that is represented by the dotted line. This denotes a small leakage which can relieve blockade, often taken as a phenomenological rate constant, such as in \cite{Levitov_2}. This slow transition rate, as we will describe in detail in the next section, is generally due to a spin selection rule that influences the respective transition rate $R_{B \rightarrow A}$. 
\\ \indent When the two states $|B\rangle$ and $|C\rangle$ are brought into resonance (Fig. \ref{Flow}(b)) by means such as external magnetic field, hyperfine interaction is activated between them which causes electrons to spin-flip between the two states at the cost of nuclear spin-flops. The relevant spin-flip from $|B\rangle$ to $|C\rangle$ is marked by a green arrow in Fig. \ref{Flow}(b). The spin-flip rate from $|C\rangle$ to $|B\rangle$ while being significant in magnitude, is of little consequence since $|C\rangle$ is a conducting state, due to which the probability of occupation of the $|C\rangle$ is very small. This system is physically described by solving self-consistently the electron transport coupled with the nuclear dynamics \cite{Inarrea}, as shown in Fig. \ref{Flow}(c). We now detail the steps shown in the flow chart. 
\\ \indent The electron transport is described by the many-body master equation approach \cite{Beenakker,Basky_Beenakker,Basky_Datta,Timm} with the hyperfine induced spin flip rates incorporated and is described in terms of the occupation probabilities $P_{N,i}$ of each $N$ electron
Fock state $|N,i\rangle$ with total energy $E^N_i$. The index $i$ here labels the states within the $N$ electron subspace. 
This equation then involves tunneling transition rates $R_{{(N,i)} \rightarrow {(N \pm 1,j)}}$ between states $|N,i \rangle$, and $|N \pm 1, j \rangle$ differing by a single electron and spin-flip transition rates $R_{{(N,i)} \rightarrow {(N,k)}}^{sf}$ between states $|N, i \rangle$, and $|N, k \rangle$ having different spin symmetries with the same number of electrons, leading to a set of equations defined by the size of the Fock space:
\begin{eqnarray}
\frac{dP_{N,i}}{dt} &=& \sum_{j}(-R_{{(N,i)}\rightarrow {(N \pm 1,j)}}P_{N,i} +  R_{{(N \pm 1, j)}\rightarrow {(N,i)}}P_{N \pm 1, j}) \nonumber \\
&+& \sum_{k} (-R_{{(N,i)}\rightarrow {(N,k)}}^{sf}P_{N,i} + R_{{(N,k)} \rightarrow {(N,i)}}^{sf}P_{N,k}),
\label{ebeenakker}
\end{eqnarray}
along with the normalization equation $\sum_{N,i}P_{N,i} = 1$. At energies close to the Fermi level, metallic contacts 
can be described using a constant density of states, parameterized using 
the bare-electron tunneling rates $\gamma_{\alpha}=\sum_{k \sigma} \frac{2 \pi}{\hbar} |t_{\alpha k \sigma, s}|^2 \delta(E-\epsilon_{k \sigma})$, where $t_{\alpha k \sigma, s}$ is the tunnel coupling matrix element with  $(\alpha=L/R)$ representing the left or right contact. 
The tunneling rates are then given by:
\begin{eqnarray}
\Gamma^{Nr}_{\alpha ij} &=& \gamma_{\alpha}|\langle N,i|\hat{d}^\dagger_{\alpha \sigma}|N-1,j\rangle|^2,  \nonumber\\
\Gamma^{Na}_{\alpha ij} &=& \gamma_{\alpha}|\langle N,i|\hat{d}_{\alpha \sigma}|N+1,j\rangle|^2,
\label{Gammas}
\end{eqnarray}
where the coherence factors $M^{Nr}_{\alpha ij}=|\langle N,i|\hat{d}^\dagger_{\alpha \sigma}|N-1,j\rangle|^2$ and $M^{Na}_{\alpha ij} =\gamma_{\alpha}|\langle N,i|\hat{d}_{\alpha \sigma}|N+1,j\rangle|^2$ represent the overlap between the many particle states involved in the tunnel transition, with ${d}^\dagger_{\alpha \sigma} (d_{\alpha \sigma})$ representing the creation (annihilation) operator of an electron in the dot which is coupled to contact labeled $\alpha = L/R$ with spin $\sigma=\uparrow/ \downarrow$ \cite{Basky_Datta}. 
The transition rates for the removal $(|N,i \rangle \rightarrow |N - 1,j \rangle)$, and addition $(|N,i \rangle \rightarrow |N+1,j \rangle)$ transitions are then given by
\begin{eqnarray}
R_{(N,i)\rightarrow(N-1,j)} &=&
\sum_{\alpha=L,R}\Gamma_{\alpha ij}^{Nr}\left[1-f \left( \frac{\epsilon^{Nr}_{ij}-\mu_{\alpha}}{k_BT_{\alpha}} \right )\right],
\nonumber\\
R_{(N,i)\rightarrow(N+1,j)} &=&
\sum_{\alpha=L,R}\Gamma_{\alpha ij}^{Na}f \left ( \frac{\epsilon^{Na}_{ij}-\mu_{\alpha}}{k_BT_{\alpha}}\right ).
\label{eq:total_rate}
\end{eqnarray}
The contact electrochemical potentials and temperatures are respectively labeled as $\mu_{\alpha}$ and $T_{\alpha}$, and $f$ is 
the corresponding Fermi-Dirac distribution function with single particle removal and addition
transport channels given by 
\begin{eqnarray}
\epsilon^{Nr}_{ij} = E^N_i - E^{N -1}_j,  \nonumber \\ 
\epsilon^{Na}_{ij} =E^{N+1}_j - E^N_i.
\label{eq:def_tc}
\end{eqnarray}
The set of transport channels ${\epsilon^{Nr/Na}_{ij}}$ and the matrix coherence factors $M^{Nr/Na}_{\alpha ij}$ serve as inputs to the master equation in \eqref{ebeenakker} that is solved self consistently with the nuclear dynamics, which involve the evaluation of the spin flip transition rates $R^{sf}$ to be described below.
\\ \indent In order to describe the spin-flip rates, we start with the Hamiltonian for spin-flip interaction, which is given by $\hat{H}_{HF} = \sum_k^{N_{nuc}}J_k \hat{{\bf{I}}}_k\cdot \hat{{\bf{S}}}$, where $\hat{{\bf{S}}}$ is the electron spin operator, $\hat{{\bf{I}}}_k$ is the nuclear spin operator and $J_k = J_{eff}\nu_0|\psi_k|^2$ is the hyperfine interaction parameter of an individual nucleus treated as a point particle. Here, $J_{eff}$ is the material specific hyerfine coupling energy, $\nu_0$ is the volume of the unit cell and $\psi_k$ the electron wavefunction at the nucleus site $k$. $N_{nuc}$ is the number of nuclei in the nuclear bath. This Hamiltonian can be expanded as
\begin{equation}\label{ham}
\hat{H}_{HF} = \sum_k^{N_{nuc}}J_k \hat{I}_k^z \hat{S}_z + \frac{1}{2}\sum_k^{N_{nuc}}J_k( \hat{I}^-_{k} \hat{S}^+ + \hat{I}^+_k \hat{S}^-)
\end{equation}
The second term above is the {\it{spin-flip}} part abbreviated hereafter as $\hat{H}_{sf}$. Under a mean field approximation, the first term may be treated as an effective magnetic field on the electrons and may be lumped with the Zeeman term due to the applied magnetic field $B$ in the electronic Hamiltonian as
\begin{equation}\label{hz}
\hat{H}_z = (g\mu_BB + J_{eff}\langle \hat{I}_z\rangle)\hat{S}_z,
\end{equation}
where $g = -0.44$ for GaAs, $\mu_B$ is the Bohr magneton and $J_{eff}$ is the total hyperfine coupling parameter summed over all nuclei and $\langle \hat{I}_z\rangle$ is the average nuclear polarization. The term $J_{eff}\langle \hat{I}_z\rangle$ relates to the Overhauser field. The spin-flip term can be treated within the Fermi's golden rule approximation in which the rate of transition from an initial state in the $|i\rangle$ to a final state $|f\rangle$ in the {\it{electron-nuclear Fock space}} is given by 
\begin{equation}\label{rsf}
R^{sf}_{i\rightarrow f} = \frac{2\pi}{\hbar}|\langle f|\hat{H}_{sf}|i\rangle|^2\rho(E)
\end{equation} 
where $\rho(E = E_f - E_i) \rightarrow \frac{\eta}{(E_f - E_i)^2 + \eta^2}$ represents the Lorentzian density of states associated with a spin-flip transition and $\eta$ is the lifetime broadening of the final state of the transition (assumed to be of the order of 0.1 $\mu$eV). Given the long time-scales of nuclear spin-relaxation compared to the electron-transport time-scales, one can decouple the fast dynamics of electron transport from the slow dynamics of the nuclei. We shall detail the specifics of nuclear dynamic equations directly in the appropriate sections. Let us now apply this to a toy model, viz. a single quantum dot coupled with a nuclear bath.
\section{Analysis of a single resonance}
\begin{figure}
	\centering
		\includegraphics[width=2.6in,height=2.5in]{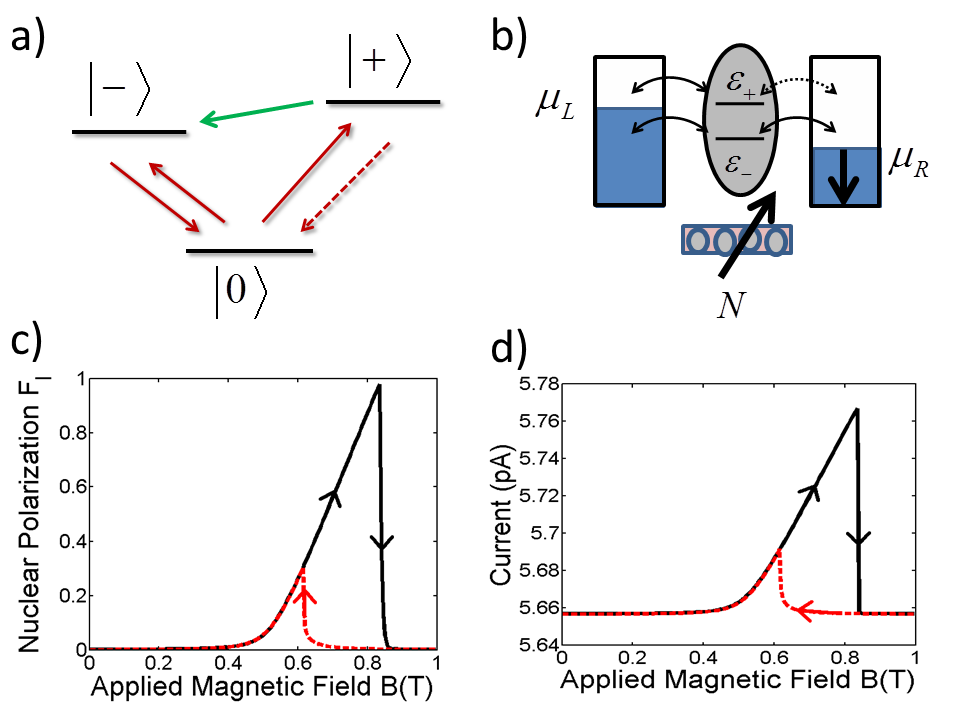}
		\caption{Single resonance hysteresis a) The Fock Space picture for the states $|0\rangle$, $|+\rangle$ and $|-\rangle$. The dotted line indicates a weak removal coupling to the one-electron state $|+\rangle$ which cannot be emptied easily. b) A physical realization of the situation in the one electron picture comprises a spin split energy level sandwiched between an unpolarized contact and a down polarized contact. The green line connecting $|+\rangle$ and $|-\rangle$ indicates a possible spin-flip interaction under resonance conditions. c) Simulated nuclear polarization $F_I$ with respect to the applied field $B$. Note the hysteresis in this case (see text) d) Simulated leakage current $I-B$ characteristic shows a similar trend.}
\label{SQD}
\end{figure}

We apply the above formulation on a toy model, viz. a spin-blockaded single quantum dot \cite{Basky_Datta} as depicted in Fig.~\ref{SQD}(b). The purpose of this discussion is to describe the physics of one resonance, which shall be extended to the double quantum dot later. There are three electronic Fock states relevant for our discussion, viz., the empty state $|0\rangle$ and two one-electron states $|+\rangle$ and $|-\rangle$ corresponding to $S_z = \pm 1/2$. The states $|+\rangle$ and $|-\rangle$ are not degenerate at $B=0$ purely to mimic the situation of energy eigenstates in the double quantum dot which is the primary objective of this paper. 
\\ \indent In this paper, the system comprises of the single orbital Anderson impurity-type 
quantum dot subject to Coulomb interaction described by the following one-dot Hubbard Hamiltonian:
\begin{equation}
\hat{H}_D=\sum_{\sigma} \epsilon_{\sigma} \hat{n}_{\sigma} + U \hat{n}_{\uparrow} \hat{n}_{\downarrow},
\label{Ham_def}
\end{equation}
where $\epsilon_{\sigma}$ represents the orbital energy, $\hat{n}_{\sigma} = \hat{d}_{\sigma}^{\dagger} \hat{d}_{\sigma}$ is the occupation number operator of an electron with 
spin $\sigma=+$, or $\sigma=-$, and $U$ is the Coulomb interaction between electrons of opposite spins occupying the same orbital. In conventional terminology, $\sigma=+$ would represent up-spins and $\sigma=-$ would represent down spins. The exact-diagonalization 
of the system Hamiltonian then results in four Fock-space energy levels labeled by their total energies 
$0,\epsilon_{+},\epsilon_{-}$ and the doubly occupied $\epsilon_{+}+\epsilon_{-} +U$ (which lies outside the transport window in our setup).   
In this section, for illustration purpose and without loss of generality, we consider $\epsilon_{+}>\epsilon_{-}$.
A physical realization of this system in the one-electron picture is shown in Fig.~\ref{SQD}(b). The spin blockade (SB) regime sets in when the applied bias is such that the $|+\rangle$ state is occupied (Fig.~\ref{SQD}(a)), thus pushing the conducting state out of the transport window, since the Coulomb interaction is large enough to prevent double occupation at any relevant bias. 
\\ \indent Spin blockade conditions may be lifted directly via a small leakage rate from $|+\rangle$ to $|0\rangle$ that relieves blockade, calculated using \eqref{eq:total_rate}. The small leakage rate in this case could arise, in principle, due to finite contact polarization and is rigorously captured in the evaluation of the matrix coherence factors defined in \eqref{Gammas}. However, when the two levels are in/near resonance, the dominant means to lift blockade is by hyperfine mediated spin-flips from $|+\rangle$ to $|-\rangle$, whose transition rates can be evaluated by summing over all possible nuclear configurations \cite{Platero_1}. Denoting the initial and final nuclear states as $|n_i\rangle$ and $|n_f\rangle$ respectively, we get 
\begin{equation}
R^{sf}_{+\rightarrow -} = \frac{2\pi}{\hbar}\sum_{n_i, n_f}|\langle -|\langle n_f| \hat{H}_{sf}|+\rangle|n_i\rangle|^2\rho(E)P(n_i)
\end{equation}
where $P(n_i)$ is the probability of the initial nuclear state being $|n_i\rangle$. This is to account for all possible configurations the nuclei could assume while the initial and final electronic states are $|+\rangle$ and $|-\rangle$ respectively. Substituting the spin-flip Hamiltonian from \eqref{ham} and noting $\langle-|\hat{S}^-|+\rangle = 1$ and $\langle-|\hat{S}^+|+\rangle = 0$, we get
\begin{equation}\label{rsf_1}
R^{sf}_{+\rightarrow -} = \frac{2\pi}{\hbar}\sum_{n_i, n_f}|\langle n_f|\frac{1}{2}\sum_k^{N_{nuc}}J_k\hat{I}^+_k| n_i\rangle|^2\rho(E)P(n_i)
\end{equation}
A non-zero matrix element requires $|n_f\rangle$ = $\hat{I}^+_k|n_i\rangle$. This implies that the k\textsuperscript{th} nucleus in $|n_i\rangle$ must be a down-spin and that in $|n_f\rangle$ must be an up-spin. Summing over all possible states of all nuclei other than the k\textsuperscript{th} yields $R^{sf}_{+\rightarrow -} = \frac{\pi}{2\hbar}\sum_{k = 1}^{N_{nuc}}|J_k|^2\rho(E)P_k(\downarrow)$, where $P_k(\downarrow)$ is the probability that the k\textsuperscript{th} nucleus is in the down-spin state. Using $\frac{N_\uparrow - N_\downarrow}{N_{nuc}} = \langle \hat{I}_z\rangle =: F_I$, and $\frac{N_\uparrow + N_\downarrow}{N_{nuc}}=1$
we get $P_k(\downarrow) = N_\downarrow/N_{nuc} = (1-F_I)/2$. Therefore, 
\begin{equation}\label{rsf_2}
R^{sf}_{+\rightarrow -} =\frac{\pi}{2\hbar}\sum_{k = 1}^{N_{nuc}}|J_k|^2\rho(E)\frac{1-F_I}{2} \approx \frac{\pi}{2\hbar}\frac{|J_{eff}|^2}{N_{nuc}}\rho(E)\frac{1 - F_I}{2}
\end{equation}
where $J_{eff} = \sum_kJ_k$ assuming $\psi_k$ to be smooth on the scale of $\nu_0^{1/3}$. The computed term is used in \eqref{ebeenakker}. An expression for $R^{sf}_{-\rightarrow+}$ may be calculated in a similar fashion. As stated before, since $|-\rangle$ is a conducting state, the probability of occupation of this state will be very small compared to the $|+\rangle$ state, and therefore the rate $R^{sf}_{-\rightarrow+}$ will have little effect on either polarization or current. Having computed all electron rates, we can obtain the dynamics of the collective nuclear polarization $F_I=\frac{N_\uparrow - N_\downarrow}{N_{nuc}}$ from the individual master equations for $N_\uparrow$ and $N_\downarrow$ as
\begin{equation}\label{fi}
\frac{dF_I}{dt} = \gamma^{nuc}\left [ (P_{+} - P_{-}) -(P_{+} + P_{-})F_I \right ] - \gamma_IF_I,
\end{equation}
where $P_{\pm}$ represent the electronic occupation probabilities, $\gamma^{nuc}=\frac{\pi}{2\hbar}\frac{|J_{eff}|^2}{N^2_{nuc}} \rho(E)$ with $N_{nuc} = 10^4$ and $\hbar\gamma_I \sim 10^{-12}$ meV is a phenomenological nuclear spin relaxation constant. The toy model can be simulated with any set of reasonable parameters $J_{eff}$, $\gamma_I$ and energies $\epsilon_{\pm}$, for unlike in the case of the DQD which we shall see next, the behavior of a single resonance is not very sensitive to the parameter set. For the simulation of Fig. \ref{SQD}(c) and (d), we have used $J_{eff}\approx 7\ \mu$eV. Typically, time-scales associated with nuclear spin relaxation are of the order of a few seconds. The transport is obtained by solving the fast electron dynamics \eqref{ebeenakker} self-consistently with the slowly varying nuclear dynamics \eqref{fi}. Finally, the steady-state solution to Eq.(\ref{ebeenakker}), set by $\frac{dP^N_i}{dt}=0$, is used to obtain the terminal current associated with contact $\alpha$:
\begin{eqnarray}
I^{\alpha} = -q\sum_{N=1}^{N_{tot}}\sum_{ij}[R^{\alpha}_{(N- 1, j)\rightarrow(N,i)}P^{N-1}_j \nonumber\\
   -R^{\alpha}_{(N,i)\rightarrow(N - 1,j)}P^N_i ],
\label{eq:curr_exp}
\end{eqnarray}
where $N_{tot}$ is the total number of electrons in the system. \newline
\indent Results of the single resonance are shown in Figs.~\ref{SQD} (c), (d). At a high enough value of $B$, the energy levels $\epsilon_{+}$ and $\epsilon_{-}$ come in resonance, thereby activating spin-flips. As the nuclear spins gradually polarize, the Overhauser field counters the decrease in energy due to $B$ given $J_{eff} > 0$ (see \eqref{hz}). We term this behavior `negative feedback', and this results in a polarization buildup which is characterized by an almost linear rise in $F_I$ versus $B$ (Fig.~\ref{SQD}(c)), until a point when the resonance breaks and $F_I$ reduces over time to $0$ due to nuclear spin-relaxation. Correspondingly, the energy of the blocking state $\epsilon_{+}$ sharply falls since the term $J_{eff}F_I$ vanishes. As a result, the level $\epsilon_{+}$ is farther below $\epsilon_{-}$ than when the resonance ended, precisely by an amount $J_{eff}F_I$ just before the resonance broke. This explains the hysteresis between the forward and reverse sweeps. On the reverse sweep, the abrupt increase in the nuclear polarization is due to positive feedback, i.e. a change in external magnetic field causes a change in the Overhauser field in the same direction. The current (Fig.~\ref{SQD}(d)) simply follows Fig.~\ref{SQD}(c) since blockade is dominantly lifted due to hyperfine-mediated spin-flips. This leads us to analyze the DQD system in which a double resonance is observed with the above explained phenomena occurring individually for each resonance, with positive and negative feedbacks juxtaposed to result in a flat-topped current. \\
\section{Analysis of a DQD double resonance} 
In the case of the DQD, we need to treat two nuclear baths with polarization variables $F_I^a$ and $F_I^b$ along with a full consideration of the electronic structure of the DQD Hamiltonian \cite{Basky_Datta}, arising from the coupling of two dots A and B (Fig.~\ref{Fig1}(b)).The DQD Hamiltonian depicted in Fig.~\ref{Fig1}(b) is described by:
\begin{eqnarray}
\hat{H}_{DQD}&=&\sum_{m} \epsilon_{m} n_{m} + \sum_{p} ( t_{mp} d_{m}^{\dagger}d_{p} + c.c ) \nonumber \\
   &+& \sum_{m} U_{m \uparrow m \downarrow}n_{m \uparrow} n_{m \downarrow} + \frac{1}{2} \sum_{p} U_{mp}n_m n_p,
\label{DQD_H}
\end{eqnarray}
where the summation indices run over the one-particle states $|a\rangle$, $|\bar{a}\rangle$ and $|b\rangle$, $|\bar{b}\rangle$ on dots A and B respectively (the bar denotes a spin-down state). The DQD parameters $\epsilon_{a/\bar{a}}$, $\epsilon_{b/\bar{b}}$ represent the on-site energies, $t_{a/\bar{a},b/\bar{b}}$ the inter-dot coupling, $U_{a,\bar{a}}=U_{b,\bar{b}}$ the on-site Coulomb matrix elements and  $U_{ab}=U_{a,\bar{a}/b,\bar{b}}/2$ the off-diagonal Coulomb matrix elements respectively.
\\ \indent Relevant to this paper is the two-electron subspace which has $6$ states with three $S = 0$ and hence $S_z = 0$ states, while the other three have $S = 1$ and hence $S_z = -1, \ 0$ and $1$. The former three are called singlets ($| S_0 \rangle$, $| S_1 \rangle$ and $ | S_2 \rangle$) and the latter, triplets ($| T_{-1} \rangle$, $| T_0 \rangle$ and $|T_{+1} \rangle$). The three triplets are degenerate at $B=0$.
\begin{figure}
	\centering
		\includegraphics[width=3.3in,height=2.7in]{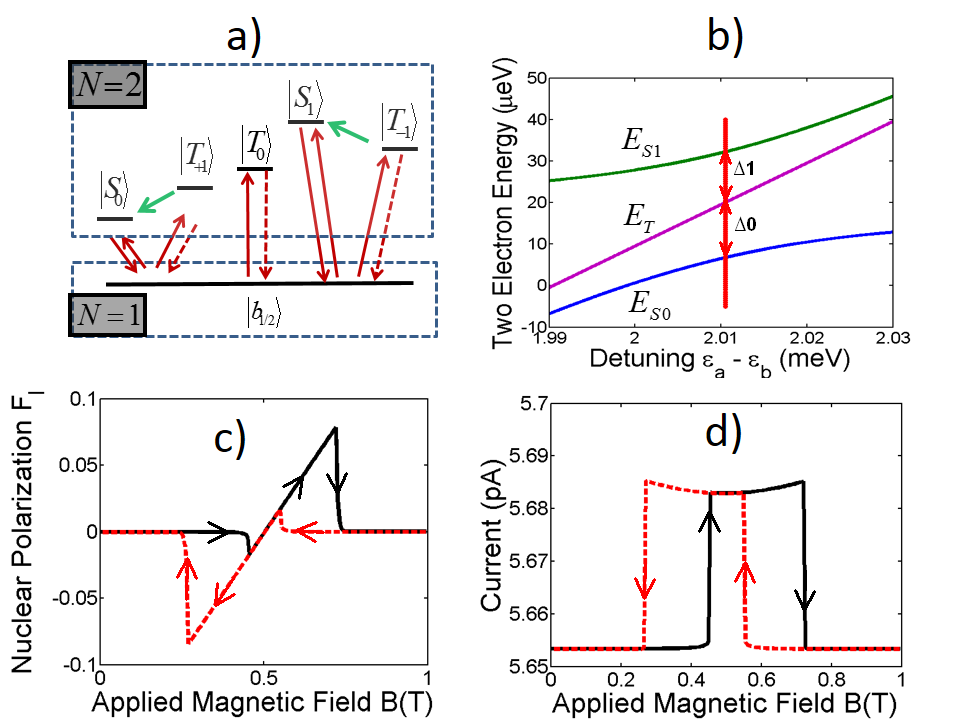}
		\caption{Double resonance hysteresis (see \cite{Levitov_2}) a) Fock state diagram of the DQD states involved in the double resonance. b) Plot of the two-electron energies with respect to the detuning between the ground state energy levels at $B=0$. The detuning is chosen so that the two resonances are energetically close to each other. c) Nuclear polarization $F_I=F_I^{a/b}$ build-up due to the double resonance and the resulting additive nature of the d) current trace with a flat topped behavior. }
\label{DQD}
\end{figure}
The relevant transport window involving $6$ states is shown in Fig.~\ref{DQD}(a), comprising the $3$ blocking triplets, $2$ conducting singlets and a spin $1/2$ bonding state $| b_{1/2} \rangle$ in the $N=1$ subspace which is accessed during the transport process. The two triplet blocking states $| T_{-1} \rangle$ and $| T_{+1} \rangle$ energetically move in opposite directions under an applied magnetic field. The two singlet states labeled $| S_0 \rangle$ and $| S_1 \rangle$ are the conducting states that would lift the blockade when mediated by hyperfine spin-flips. The DQD parameters $\epsilon_{a}=0.01$ meV, $\epsilon_{b}=-2$ meV, $t=9$ $\mu$eV, $U_{a,\bar{a}}=U_{b,\bar{b}}\approx 4.02$ meV, $U_{ab}=U_{a,\bar{a}}/2$, are chosen such that the system is in spin blockade \cite{Basky_Datta,Tarucha_1}. This also ensures that the two resonances between $| S_{1} \rangle$ and $| T_{-1} \rangle$ and between $| S_{0} \rangle$ and $| T_{+1} \rangle$ occur within a spacing of $\Delta_0-\Delta_1 \sim 1 \mu$eV, with $\Delta_0=|E_{S_0}-E_{T_{+1}}|$ and $\Delta_1=|E_{S_1}-E_{T_{-1}}|$, as shown in Fig.~\ref{DQD}(b). It is first important to note that the electronic structure of the two singlet states are typically given by:
\begin{equation}
| S_{0/1} \rangle = \alpha_{0/1} | a \bar{b}\rangle +\beta_{0/1} | \bar{a} b \rangle + \delta_{0/1} | a \bar{a} \rangle + \xi_{0/1} | b \bar{b} \rangle,
\label{singlet}
\end{equation} 
where $\alpha_{0/1}$,  $\beta_{0/1}$, $\delta_{0/1}$ and $\xi_{0/1}$ represent various wavefunction coefficients of $| S_{0/1} \rangle$ states which represent the hybridization among the respective two particle basis states. It should be noted that for the aforementioned parameter set, the ground state singlet permits significant hybridization between states that have two electrons: those with one electron each on either dot and a state with double occupation on one of the dots (dot B in this case). The doubly occupied state on dot A is almost absent i.e. $\delta_{0/1}\approx 0$ since it is energetically much higher than the other basis states. The remaining coefficients are significant and play a role in either spin-flip, or in coupling to contacts. As clearly pointed out in Ref. \cite{Basky_Datta}, this is crucial in making the state $|S_0 \rangle$ a conducting state. Furthermore, the different extents of hybridizations between the two singlet states that are defined by the above coefficients will play a crucial role in the asymmetric aspects of the hysteresis to be explained below.

Blockade may be relieved by a small leakage rate from each of the triplets to the one-electron bonding states calculated via \eqref{eq:total_rate}, but predominantly via the spin-flip transition rates between $| T_{+1} \rangle$ and $| S_0 \rangle$ and between $| T_{-1} \rangle$ and $| S_1 \rangle$. The computation of these rates have several subtleties, which shall be detailed now. \\
\indent The spin-flip Hamiltonian $\hat{H}_{sf}$ from \eqref{ham} now has contributions from nuclei of both dots, which can be written as
\begin{equation}\label{bighsf}
\hat{H}_{sf} = \frac{1}{2}\sum_{k\in A}^{N_a}J_k( \hat{I}^-_{k} \hat{S}^+ + \hat{I}^+_k \hat{S}^-) + \frac{1}{2}\sum_{k\in B}^{N_b}J_k( \hat{I}^-_{k} \hat{S}^+ + \hat{I}^+_k \hat{S}^-) 
\end{equation}
The electron-nuclear Fock space states involved in the golden rule calculation are now of the form $|i\rangle = | T_{+1}\rangle \otimes | n^A_i\rangle \otimes | n^B_i\rangle$, $|f\rangle =| S_{0}\rangle \otimes | n^A_f\rangle \otimes | n^B_f\rangle$ (similarly for the other resonance involving $| T_{-1} \rangle$ and $| S_1 \rangle$). We must therefore sum over all possible nuclear configurations of dots A and B in order to calculate the spin-flip rates between the electronic states $ | T_{\pm 1} \rangle$ and $|S_{0/1}\rangle$ via Fermi's Golden Rule. If the final nuclear state is such that a nucleus $j$ on dot A has flopped in exchange for the electronic spin flip with the rest of the nuclei have retained their state, i.e., $|n^A_f\rangle = \hat{I}_j^+|n^A_i\rangle$ and $|n^B_f\rangle = |n^B_i\rangle$ , then for the \textit{second} term in \eqref{bighsf} 
\begin{equation}\label{zero}
\langle f|\sum_{k\in B}^{N_b}J_k( \hat{I}^-_{k} \hat{S}^+ + \hat{I}^+_k \hat{S}^-)|i\rangle = 0
\end{equation}
since all nuclei on dot B have retained their original state and thus the matrix element $\langle n^B_f|I_k^{\pm}|n^B_i\rangle = \langle n^B_i|I_k^{\pm}|n^B_i\rangle =0$ for $\ k\in B$. Thus a spin-flip on dot A will make the contribution from the second term equal to 0. Similarly, a spin-flip on dot B will make the contribution from the first term of \eqref{bighsf} equal to 0. Of course, if the final and initial nuclear states are such that more than one of the nuclei have flipped their states, then contributions from \textit{both} terms will be 0. Therefore, for a non-zero contribution from the first term, we require $|n^B_f\rangle = |n_i^B\rangle$, but this makes the contribution from the second term zero as per \eqref{zero}, while the opposite is the requirement for a non-zero contribution from the second term, which similarly makes the contribution from the first term zero. Thus, we see that in summing over all possible nuclear configurations, the first and second terms in $\hat{H}_{sf}$ above can \textit{never contribute together}. This leads to two separate terms, each of the form \eqref{rsf_1}, one by tracing over all dot B nuclei and then calculating the matrix element over dot A, and the other by tracing over dot A and then calculating the matrix element over dot B. Each of these terms finally assume a form similar to \eqref{rsf_2}. In addition, we also note that $\langle f|\hat{H}_{sf}|i\rangle$ consists of a matrix element between the electronic states based on \eqref{singlet} which is \textit{dynamically dependent} on the values of $B$ and $F_I^{a/b}$. This therefore requires us to also compute the electronic matrix elements self-consistently with electronic transport and nuclear polarization. We have  $J_{eff} = \sum_{k\in A}J_k = \sum_{k\in B}J_k$, which is a material specific constant. Therefore, 
\begin{eqnarray}\label{rsf_two}
R^{sf}_{T\rightarrow S} &=& \frac{\pi}{2\hbar}(\sum_k^{N_a}|J_k\beta_0|^2\frac{1-F_I^a}{2} + \sum_k^{N_b}|J_k\alpha_0|^2\frac{1-F_I^b}{2})\rho(E) \nonumber \\
&\approx& \frac{\pi}{2\hbar}(\frac{J_{eff}^2\beta_0^2}{N_a}\frac{1-F_I^a}{2} + \frac{J_{eff}^2\alpha_0^2}{N_b}\frac{1-F_I^b}{2})\rho(E),
\end{eqnarray}
where $\alpha_0$ and $\beta_0$ are from \eqref{singlet} and $ T $ represents the $| T_{+1} \rangle$ state and $S$ represents the $ | S_{0} \rangle$ state. $N_a$ and $N_b$ are respectively the number of nuclei on dots A and B. The transition rates between the other pairs involved in the second resonance can be similarly derived. The spin-flip rates in the reverse direction viz. from the singlets to the triplets while being equal to \eqref{rsf_two} as well (with the $1-F_I^{a/b}$ factors replaced by $1+F_I^{a/b}$), can be neglected in this problem since the singlets being conducting states have very low occupation probabilities compared to the triplets and therefore the reverse spin-flip rate has little effect on either current or polarization. It is essential to note that we have an incoherent contribution from the two baths, i.e., there are no cross-terms linking contributions from both baths. This is simply because the electronic eigenstates are delocalized over the dots, while the nuclear eigenstates are localized on each dots. Eq. \eqref{rsf_two} shows us that there will be a non-zero spin-flip rate whether or not the hyperfine coupling parameters are equal and hence the matrix element of the spin-flip Hamiltonian between the singlet and triplet \textbf{does not} go to 0 even if the baths have equal Overhauser fields. This can be confirmed from \cite{Levitov_3} where similar expressions were employed for spin-flip rates, since Eq. (3-4) in \cite{Levitov_3}, whose sum gives the flip-rate from a triplet to the $S_z = 0$ subspace, do not add up to zero for they also incorporate two polarization variables. \\
\indent Finally, it is also important to note that in the scenario of equal Overhauser fields, the state $| T_0 \rangle$ does not affect the nuclear polarizations. This is because $| T_0 \rangle$ can spin-flip only to $| T_+ \rangle$ or $|T_- \rangle$, which at all values of $B$ are symmetrically placed w.r.t. $| T_0 \rangle$ in the energy space. This implies that the spin-flip rates into $| T_0 \rangle$ from $| T_+ \rangle$ and $| T_- \rangle$ and vice-versa are \textit{always equal}. Since the change in nuclear spin due to a transition between $| T_0 \rangle$ and $| T_+ \rangle$ is exactly the opposite of the change caused by a transition between $| T_0 \rangle$ and $| T_- \rangle$, these effects perfectly cancel, making the state $| T_0 \rangle$ incapable of affecting nuclear dynamics for the case of equal Overhauser fields. Furthermore, if $B$ is sufficiently large, the triplets $| T_{\pm 1} \rangle$ are no longer in resonance with $| T_0 \rangle$, and the spin-flip rate becomes negligible. We shall however see that $| T_0 \rangle$ occupies an important role, albeit for a different reason, when the Overhauser fields are not equal. \\
\indent Let us define $\gamma^a_1 = \frac{\pi}{2\hbar}|\frac{J_{eff}\beta_0}{N_a}|^2\rho(E_1)$, $\gamma^a_2 = \frac{\pi}{2\hbar}|\frac{J_{eff}\alpha_1}{N_a}|^2\rho(E_2)$, $\gamma^b_1 = \frac{\pi}{2\hbar}|\frac{J_{eff}\alpha_0}{N_b}|^2\rho(E_1)$ and $\gamma^b_2 = \frac{\pi}{2\hbar}|\frac{J_{eff}\beta_1}{N_b}|^2\rho(E_2)$, where $E_1 = E_{T_{+1}} - E_{S_0}$ and $E_2 = E_{S_1} - E_{T_{-1}}$. The dynamics of $F_I^{a/b}$ are then governed by an equation similar to \eqref{fi}:
\begin{eqnarray*}\label{bigfi}
\begin{split}
\frac{dF_I^{a/b}}{dt} = -[\gamma^{a/b}_1(P_{S_0} + P_{T_{+1}}) + \gamma^{a/b}_2(P_{S_1} + P_{T_{-1}})]F_I^{a/b} + \\ [\gamma^{a/b}_1(P_{T_{+1}} - P_{S_0}) + \gamma^{a/b}_2(P_{S_1} - P_{T_{-1}})] - \gamma_IF_I^{a/b}
\end{split}
\end{eqnarray*}
Let us first consider the case where $N_a = N_b$. Without the inclusion of statistical fluctuations in the nuclear field, the dynamics of $F_I^a$ and $F_I^b$ are identical and consequently there is no difference Overhauser field. For transport dynamics, we choose the coupling to contacts as $\hbar\gamma=5.2 \times 10^{-3}$ meV, the nuclear coupling parameter as $J_{eff}=70$ $\mu$eV \cite{Fischer}, the nuclear spin-relaxation constant as $\hbar\gamma_I = 10^{-12}$ meV and the number of nuclei $N_a = N_b = 10^5$. In the setup of Fig. \ref{Fig1}(a), coupling to contacts ought to be smaller than the peak spin-flip rate. This is because while the spin-flip rate increases by several orders of magnitude as the energy levels attain resonance, the current rises only by a fraction of a picoampere, thereby necessitating the coupling to contacts to be the rate-limiting factor for current at resonance. The characteristic of any one of $F_I^a$ and $F_I^b$ versus $B$ is shown in Fig.~\ref{DQD}(c). We note that this consists of two dragged resonances, i.e. two polarization curves each resembling Fig. \ref{SQD}(c). With the chosen DQD parameter set, in the forward $B$ sweep the first energetically feasible resonance is that of $| T_{-1} \rangle- |S_1 \rangle$ which has a positive feedback effect between the applied field $B$ and the Overhauser field. The $| T_{+1} \rangle- | S_{0} \rangle$ resonance energetically follows almost simultaneously with a negative feedback effect. The polarization arising out of $| T_{-1} \rangle- |S_1 \rangle$ transitions is flipped with respect to that arising out of $| T_{+1} \rangle-| S_0 \rangle$ transitions since the former requires spin-raising while the latter requires spin-lowering. Nevertheless, both these resonances individually produce a current waveform similar to Fig.~\ref{SQD}(d), albeit laterally flipped with respect to one another. Thus the two \textit{triangular} current waveforms superpose to result in the flat-topped \textit{square} waveform as depicted in Fig.~\ref{DQD}(d). It is important to note that the flat-top is observed because the two resonances occur sufficiently close to one another.\\
\begin{figure}
	\centering
		\includegraphics[width=3.3in,height=2.6in]{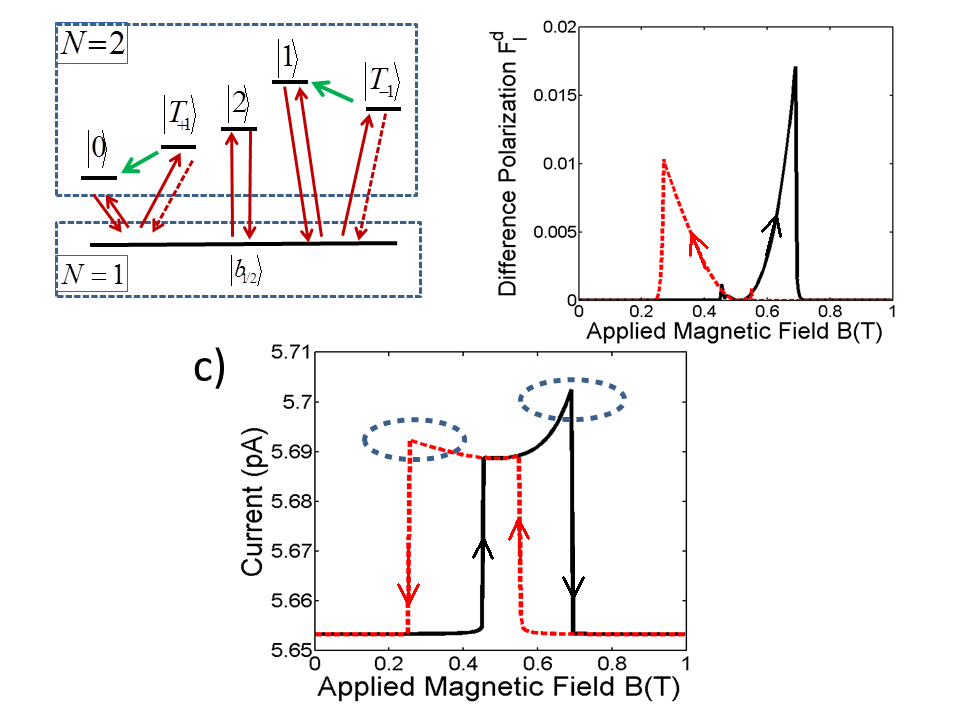}
		\caption{Effect of the difference Overhauser field a) New Fock space states arising out of mixing of the $S_z = 0$ subspace when difference field is non-zero. b) Difference Overhauser polarization $F_I^d=|F_I^a-F_I^b|$ that opens up transport channels resulting in c) Flare up of currents similar to what is noted in the experiments in \cite{Ono}. The asymmetry of this fin-structure is due to different Overlap elements of the spin-flip Hamiltonian with structurally different singlet states involved in the double resonance.}
\label{Diff_Field}
\end{figure}
\indent The second feature of interest is the fin-like flare-up that we observe towards the ends of the resonances, encircled in Fig.~\ref{Fig1}(a). We attribute this to two possible phenomena. The first one relates to the two resonances whose energetic spacing $\Delta_0-\Delta_1$ relies on the DQD structure. The two resonances may be pushed farther apart due to a different parameter set, leading to an imperfect superposition of the two triangular current waveforms. This can be readily visualized to cause fin-like rise towards the end and a caving-in in the middle, which results in an overall decrease in the current of the middle portion, as is also observed in the current trace in \cite{Levitov_2}. 
\\ \indent The second one is more subtle and indicates the presence of a difference Overhauser field. This can arise due to fluctuations in the Overhauser fields \cite{Taylor,Lukin,SB_1} of the two dots and/or due to different dynamics of the polarization variables. The latter situation may arise, for example, due to unequal sizes of the dots, leading to $N_a \neq N_b$. We shall focus on this scenario. In this case, the dynamics of $F_I^a$ and $F_I^b$ are in general different, producing a non-zero difference Overhauser field. An important consequence of the difference Overhauser field is that it mixes all states in the $S_z = 0$ subspace of the Hamiltonian, making the new eigenstates of the electronic Hamiltonian $| T_{\pm 1} \rangle$ and the four other states, each of which is a linear combination of $|S_0 \rangle, |S_1 \rangle, |S_2 \rangle$ and $| T_0 \rangle$. We now annotate the three $S_z = 0$ eigenstates in the relevant transport window as $|0\rangle$, $|1\rangle$ and $|2\rangle$. As depicted in Fig. \ref{Diff_Field}(a), state $|0\rangle$ takes the place of $| S_0 \rangle$, $|1\rangle$ takes the place of $| S_1 \rangle$ and $|2\rangle$ takes the place of $| T_0 \rangle$. We then obtain two blocking states ($| T_{\pm 1} \rangle$) and {\it{three}} conducting states in our transport window since the state that was purely a $| T_0 \rangle$ previously is now replaced by $|2\rangle$, a linear combination of the conducting singlets and $| T_0 \rangle$. Therefore, the dotted arrow from $| T_0 \rangle$ to $|b_{1/2}\rangle$ in Fig. \ref{DQD}(a) which represented a weak leakage rate is now replaced by a thick arrow from $|2\rangle$ to $|b_{1/2}\rangle$ in Fig. \ref{Diff_Field}(a), representing a conducting state. Thus, whenever there is a build-up of a difference Overhauser field, there is a rise in the current due to the additional conducting state. In Fig.~\ref{Diff_Field}(b), we plot the difference Overhauser polarization versus $B$. One can immediately note that in the regions where there is a build-up of the difference polarization, there is a fin-like rise in current (Fig.~\ref{Diff_Field}(c)). The polarization $F_I^{a/b}$ representing the sum Overhauser field on the other hand remains sufficiently similar to Fig.~\ref{DQD}(b) and is hence omitted.
\\ \indent A curious feature of Fig. \ref{Diff_Field}(c) is the stronger peak in the current on the right side as compared against the left. For this, we first note that at zero difference field, the (0, 2) component, i.e., $|b\bar{b}\rangle$, has the lowest energy for the given parameter set and therefore, the lower eigenstate $|S_0\rangle$ has a larger $|b\bar{b}\rangle$ component than $|S_1\rangle$. In other words, $|\xi_0| > |\xi_1|$. As Overhauser field builds up, each of $|0\rangle$ and $|1\rangle$ (and $|2\rangle$) take the general form \eqref{singlet} with new coefficients. However, the $|b\bar{b}\rangle$ coefficient of $|0\rangle$ continues to be larger than that of $|1\rangle$ since the former is energetically lower. Since the tunneling from the two-electron states to the one-electron state $|b_{1/2}\rangle$ is proportional to the square of this coefficient in accordance with \eqref{Gammas}, the rate of transition from the singlet $| S_0 \rangle$ to $|b_{1/2}\rangle$ is larger than that from to $| S_1 \rangle$ to $|b_{1/2}\rangle$. Since the rate corresponding to coupling to the contacts is smaller than the peak spin-flip rate, this implies that the number of spin-flips achieved at/near resonance depends purely on the coupling to contacts. Therefore the resonance between $| T_{+1} \rangle$ and $| S_0 \rangle$ allows more spin-flips than the resonance between $| T_{-1} \rangle$ and $| S_1 \rangle$. Consequently, in the case of differing dynamics for $F_I^a$ and $F_I^b$, the resonance between $| T_{+1} \rangle$ and $| S_0 \rangle$ enhances the difference polarization more by allowing for more spin-flips than the $|T_{-1} \rangle-| S_1 \rangle$ resonance. Since the right side of the current waveform is governed by the feedback of the $| T_{+1} \rangle-| S_0 \rangle$ resonance, we see an enhanced difference polarization towards the right, as depicted in Fig. \ref{Diff_Field}(b). Finally, since greater difference polarization implies a greater hybridization of $| T_0 \rangle$ and the singlets, the state $|2\rangle$ becomes more conducting, leading to greater current measured through the device. This explains the asymmetric peaks in the current waveform in the case of unequal Overhauser fields. It is likely that the peak that is observed on the right side of the experimental curve Fig. \ref{Fig1}(a) arises out of the presence of a difference field. The parameters used for generating the difference Overhauser field are $N_a = 10^5, N_b = 1.4\times10^5$ and $J_{eff} = 70 \ \mu$eV. It is of interest to understand here that the right-sided asymmetry in current \textit{remains unchanged} even if we reverse the sizes of the dots, i.e. $N_a > N_b$, since it originates from the inherent asymmetry in the electronic structure and not from the nuclear dynamics. This is relevant for the right-sided asymmetry also observed in the experimental current waveform Fig. \ref{Fig1}(a). Of course, a smaller difference between the number of nuclei on the two dots would produce relatively smaller peaks, but the pronounced asymmetry remains since it originates from the electronic structure of the singlets.  \\
\indent Thus, the sharp rise/`switching'  of the current is a consequence of the sum Overhauser field, while the fin-like flare-up is the effect of difference Overhauser field and/or a larger than ideal gap between the two resonances. Both the difference Overhauser field and a non-ideal resonance gap are simultaneously present in a fabricated DQD, as imperfections in deposition are bound to make the two dots unequally sized and hence the effective coupling different, and simultaneously the values of Coulomb repulsion and/or tunnel coupling strength different from the parameter set stated earlier. \\
\section{Summary} In this paper, we studied the nuances of a two-bath model with two polarization variables in the context of electronic transport and the experimentally observed bistability \cite{Ono}. In doing so, we detailed the derivation of the spin-flip rates as well as the interplay between the difference Overhauser field and the DQD electronic structure and their effects on the current bistability. However, the explanation of the unstable region in Fig.~\ref{Fig1}(a) and the associated self-sustaining current oscillations may involve a nutation in the electron nuclear space and might necessitate the use of the density matrix approach \cite{Koenig,Basky_Milena}. This aspect still remains elusive with a couple of recently proposed candidates \cite{Osc1,Osc2}. Developing an understanding of non-equilibrium situations that involves the coupling with dynamics of additional baths will form a new and important frontier in the area of nanoscale transport. \\

Acknowledgments: The authors gratefully acknowledge insightful discussions with Prof. Supriyo Datta and suggestions from Prof. Dipan Ghosh. The work in part was funded by the Department of Science and Technology India under the SERB program.

\bibliographystyle{apsrev}	
\bibliography{PAPER_OV_BIB}		

\begin{thebibliography}{27}
\expandafter\ifx\csname natexlab\endcsname\relax\def\natexlab#1{#1}\fi
\expandafter\ifx\csname bibnamefont\endcsname\relax
  \def\bibnamefont#1{#1}\fi
\expandafter\ifx\csname bibfnamefont\endcsname\relax
  \def\bibfnamefont#1{#1}\fi
\expandafter\ifx\csname citenamefont\endcsname\relax
  \def\citenamefont#1{#1}\fi
\expandafter\ifx\csname url\endcsname\relax
  \def\url#1{\texttt{#1}}\fi
\expandafter\ifx\csname urlprefix\endcsname\relax\def\urlprefix{URL }\fi
\providecommand{\bibinfo}[2]{#2}
\providecommand{\eprint}[2][]{\url{#2}}

\bibitem[{\citenamefont{Ono et~al.}(2002)\citenamefont{Ono, Austing, Tokura,
  and Tarucha}}]{Tarucha_1}
\bibinfo{author}{\bibfnamefont{K.}~\bibnamefont{Ono}},
  \bibinfo{author}{\bibfnamefont{D.~G.} \bibnamefont{Austing}},
  \bibinfo{author}{\bibfnamefont{Y.}~\bibnamefont{Tokura}}, \bibnamefont{and}
  \bibinfo{author}{\bibfnamefont{S.}~\bibnamefont{Tarucha}},
  \bibinfo{journal}{Science} \textbf{\bibinfo{volume}{297}},
  \bibinfo{pages}{1313} (\bibinfo{year}{2002}).

\bibitem[{\citenamefont{Hanson et~al.}(2007)\citenamefont{Hanson, Kouwenhoven,
  Petta, Tarucha, and Vandersypen}}]{Leo_review}
\bibinfo{author}{\bibfnamefont{R.}~\bibnamefont{Hanson}},
  \bibinfo{author}{\bibfnamefont{L.~P.} \bibnamefont{Kouwenhoven}},
  \bibinfo{author}{\bibfnamefont{J.~R.} \bibnamefont{Petta}},
  \bibinfo{author}{\bibfnamefont{S.}~\bibnamefont{Tarucha}}, \bibnamefont{and}
  \bibinfo{author}{\bibfnamefont{L.~M.~K.} \bibnamefont{Vandersypen}},
  \bibinfo{journal}{Rev. Mod. Phys.} \textbf{\bibinfo{volume}{79}},
  \bibinfo{pages}{1217} (\bibinfo{year}{2007}).

\bibitem[{\citenamefont{Muralidharan and Datta}(2007)}]{Basky_Datta}
\bibinfo{author}{\bibfnamefont{B.}~\bibnamefont{Muralidharan}}
  \bibnamefont{and} \bibinfo{author}{\bibfnamefont{S.}~\bibnamefont{Datta}},
  \bibinfo{journal}{Phys. Rev. B} \textbf{\bibinfo{volume}{76}},
  \bibinfo{pages}{035432} (\bibinfo{year}{2007}).

\bibitem[{\citenamefont{Ono and Tarucha}(2004)}]{Ono}
\bibinfo{author}{\bibfnamefont{K.}~\bibnamefont{Ono}} \bibnamefont{and}
  \bibinfo{author}{\bibfnamefont{S.}~\bibnamefont{Tarucha}},
  \bibinfo{journal}{Phys. Rev. Lett.} \textbf{\bibinfo{volume}{92}},
  \bibinfo{pages}{256803} (\bibinfo{year}{2004}).

\bibitem[{\citenamefont{Petta et~al.}(2005)\citenamefont{Petta, Johnson,
  Taylor, Laird, Yacoby, Lukin, Marcus, Hanson, and Gossard}}]{SB_1}
\bibinfo{author}{\bibfnamefont{J.~R.} \bibnamefont{Petta}},
  \bibinfo{author}{\bibfnamefont{A.~C.} \bibnamefont{Johnson}},
  \bibinfo{author}{\bibfnamefont{J.~M.} \bibnamefont{Taylor}},
  \bibinfo{author}{\bibfnamefont{E.~A.} \bibnamefont{Laird}},
  \bibinfo{author}{\bibfnamefont{A.}~\bibnamefont{Yacoby}},
  \bibinfo{author}{\bibfnamefont{M.~D.} \bibnamefont{Lukin}},
  \bibinfo{author}{\bibfnamefont{C.~M.} \bibnamefont{Marcus}},
  \bibinfo{author}{\bibfnamefont{M.~P.} \bibnamefont{Hanson}},
  \bibnamefont{and} \bibinfo{author}{\bibfnamefont{A.~C.}
  \bibnamefont{Gossard}}, \bibinfo{journal}{Science}
  \textbf{\bibinfo{volume}{309}}, \bibinfo{pages}{2180} (\bibinfo{year}{2005}).

\bibitem[{\citenamefont{Koppens et~al.}(2005)\citenamefont{Koppens, Folk,
  Elzerman, Hanson, van Beveren, Vink, Tranitz, Wegscheider, Kouwenhoven, and
  Vandersypen}}]{SB_2}
\bibinfo{author}{\bibfnamefont{F.~H.~L.} \bibnamefont{Koppens}},
  \bibinfo{author}{\bibfnamefont{J.~A.} \bibnamefont{Folk}},
  \bibinfo{author}{\bibfnamefont{J.~M.} \bibnamefont{Elzerman}},
  \bibinfo{author}{\bibfnamefont{R.}~\bibnamefont{Hanson}},
  \bibinfo{author}{\bibfnamefont{L.~H.~W.} \bibnamefont{van Beveren}},
  \bibinfo{author}{\bibfnamefont{I.~T.} \bibnamefont{Vink}},
  \bibinfo{author}{\bibfnamefont{H.~P.} \bibnamefont{Tranitz}},
  \bibinfo{author}{\bibfnamefont{W.}~\bibnamefont{Wegscheider}},
  \bibinfo{author}{\bibfnamefont{L.~P.} \bibnamefont{Kouwenhoven}},
  \bibnamefont{and} \bibinfo{author}{\bibfnamefont{L.~M.~K.}
  \bibnamefont{Vandersypen}}, \bibinfo{journal}{Science}
  \textbf{\bibinfo{volume}{309}}, \bibinfo{pages}{1346} (\bibinfo{year}{2005}).

\bibitem[{\citenamefont{Nowack et~al.}(2007)\citenamefont{Nowack, Koppens,
  Nazarov, and Vandersypen}}]{SB_3}
\bibinfo{author}{\bibfnamefont{K.~C.} \bibnamefont{Nowack}},
  \bibinfo{author}{\bibfnamefont{F.~H.~L.} \bibnamefont{Koppens}},
  \bibinfo{author}{\bibfnamefont{Y.~V.} \bibnamefont{Nazarov}},
  \bibnamefont{and} \bibinfo{author}{\bibfnamefont{L.~M.~K.}
  \bibnamefont{Vandersypen}}, \bibinfo{journal}{Science}
  \textbf{\bibinfo{volume}{318}}, \bibinfo{pages}{1430} (\bibinfo{year}{2007}).

\bibitem[{\citenamefont{Bluhm et~al.}(2010)\citenamefont{Bluhm, Foletti,
  Mahalu, Umansky, and Yacoby}}]{Yakoby}
\bibinfo{author}{\bibfnamefont{H.}~\bibnamefont{Bluhm}},
  \bibinfo{author}{\bibfnamefont{S.}~\bibnamefont{Foletti}},
  \bibinfo{author}{\bibfnamefont{D.}~\bibnamefont{Mahalu}},
  \bibinfo{author}{\bibfnamefont{V.}~\bibnamefont{Umansky}}, \bibnamefont{and}
  \bibinfo{author}{\bibfnamefont{A.}~\bibnamefont{Yacoby}},
  \bibinfo{journal}{Phys. Rev. Lett.} \textbf{\bibinfo{volume}{105}},
  \bibinfo{pages}{216803} (\bibinfo{year}{2010}).

\bibitem[{\citenamefont{Rudner and Levitov}(2007)}]{Levitov_1}
\bibinfo{author}{\bibfnamefont{M.~S.} \bibnamefont{Rudner}} \bibnamefont{and}
  \bibinfo{author}{\bibfnamefont{L.~S.} \bibnamefont{Levitov}},
  \bibinfo{journal}{Phys. Rev. Lett.} \textbf{\bibinfo{volume}{99}},
  \bibinfo{pages}{036602} (\bibinfo{year}{2007}).

\bibitem[{\citenamefont{Rudner and Levitov}(2010)}]{Levitov_2}
\bibinfo{author}{\bibfnamefont{M.~S.} \bibnamefont{Rudner}} \bibnamefont{and}
  \bibinfo{author}{\bibfnamefont{L.~S.} \bibnamefont{Levitov}},
  \bibinfo{journal}{Nanotechnology} \textbf{\bibinfo{volume}{21}},
  \bibinfo{pages}{274016} (\bibinfo{year}{2010}).

\bibitem[{\citenamefont{Rudner et~al.}(2011)\citenamefont{Rudner, Koppens,
  Folk, Vandersypen, and Levitov}}]{Levitov_3}
\bibinfo{author}{\bibfnamefont{M.~S.} \bibnamefont{Rudner}},
  \bibinfo{author}{\bibfnamefont{F.~H.~L.} \bibnamefont{Koppens}},
  \bibinfo{author}{\bibfnamefont{J.~A.} \bibnamefont{Folk}},
  \bibinfo{author}{\bibfnamefont{L.~M.~K.} \bibnamefont{Vandersypen}},
  \bibnamefont{and} \bibinfo{author}{\bibfnamefont{L.~S.}
  \bibnamefont{Levitov}}, \bibinfo{journal}{Phys. Rev. B}
  \textbf{\bibinfo{volume}{84}}, \bibinfo{pages}{075339}
  (\bibinfo{year}{2011}).

\bibitem[{\citenamefont{Jouravlev and Nazarov}(2006)}]{Nazarov}
\bibinfo{author}{\bibfnamefont{O.~N.} \bibnamefont{Jouravlev}}
  \bibnamefont{and} \bibinfo{author}{\bibfnamefont{Y.~V.}
  \bibnamefont{Nazarov}}, \bibinfo{journal}{Phys. Rev. Lett.}
  \textbf{\bibinfo{volume}{96}}, \bibinfo{pages}{176804}
  (\bibinfo{year}{2006}).

\bibitem[{\citenamefont{Danon}(2013)}]{Danon}
\bibinfo{author}{\bibfnamefont{J.}~\bibnamefont{Danon}},
  \bibinfo{journal}{Phys. Rev. B} \textbf{\bibinfo{volume}{88}},
  \bibinfo{pages}{075306} (\bibinfo{year}{2013}).

\bibitem[{\citenamefont{I\~narrea et~al.}(2007)\citenamefont{I\~narrea,
  Platero, and MacDonald}}]{Inarrea}
\bibinfo{author}{\bibfnamefont{J.}~\bibnamefont{I\~narrea}},
  \bibinfo{author}{\bibfnamefont{G.}~\bibnamefont{Platero}}, \bibnamefont{and}
  \bibinfo{author}{\bibfnamefont{A.~H.} \bibnamefont{MacDonald}},
  \bibinfo{journal}{Phys. Rev. B} \textbf{\bibinfo{volume}{76}},
  \bibinfo{pages}{085329} (\bibinfo{year}{2007}).

\bibitem[{\citenamefont{Taylor et~al.}(2007)\citenamefont{Taylor, Petta,
  Johnson, Yacoby, Marcus, and Lukin}}]{Taylor}
\bibinfo{author}{\bibfnamefont{J.~M.} \bibnamefont{Taylor}},
  \bibinfo{author}{\bibfnamefont{J.~R.} \bibnamefont{Petta}},
  \bibinfo{author}{\bibfnamefont{A.~C.} \bibnamefont{Johnson}},
  \bibinfo{author}{\bibfnamefont{A.}~\bibnamefont{Yacoby}},
  \bibinfo{author}{\bibfnamefont{C.~M.} \bibnamefont{Marcus}},
  \bibnamefont{and} \bibinfo{author}{\bibfnamefont{M.~D.} \bibnamefont{Lukin}},
  \bibinfo{journal}{Phys. Rev. B} \textbf{\bibinfo{volume}{76}},
  \bibinfo{pages}{035315} (\bibinfo{year}{2007}).

\bibitem[{\citenamefont{Gullans et~al.}(2010)\citenamefont{Gullans, Krich,
  Taylor, Bluhm, Halperin, Marcus, Stopa, Yacoby, and Lukin}}]{Lukin}
\bibinfo{author}{\bibfnamefont{M.}~\bibnamefont{Gullans}},
  \bibinfo{author}{\bibfnamefont{J.~J.} \bibnamefont{Krich}},
  \bibinfo{author}{\bibfnamefont{J.~M.} \bibnamefont{Taylor}},
  \bibinfo{author}{\bibfnamefont{H.}~\bibnamefont{Bluhm}},
  \bibinfo{author}{\bibfnamefont{B.~I.} \bibnamefont{Halperin}},
  \bibinfo{author}{\bibfnamefont{C.~M.} \bibnamefont{Marcus}},
  \bibinfo{author}{\bibfnamefont{M.}~\bibnamefont{Stopa}},
  \bibinfo{author}{\bibfnamefont{A.}~\bibnamefont{Yacoby}}, \bibnamefont{and}
  \bibinfo{author}{\bibfnamefont{M.~D.} \bibnamefont{Lukin}},
  \bibinfo{journal}{Phys. Rev. Lett.} \textbf{\bibinfo{volume}{104}},
  \bibinfo{pages}{226807} (\bibinfo{year}{2010}).

\bibitem[{\citenamefont{Lopez-Monis et~al.}(2011)\citenamefont{Lopez-Monis,
  Inarrea, and Platero}}]{Platero_1}
\bibinfo{author}{\bibfnamefont{C.}~\bibnamefont{Lopez-Monis}},
  \bibinfo{author}{\bibfnamefont{J.}~\bibnamefont{Inarrea}}, \bibnamefont{and}
  \bibinfo{author}{\bibfnamefont{G.}~\bibnamefont{Platero}},
  \bibinfo{journal}{New Journal of Physics} \textbf{\bibinfo{volume}{13}},
  \bibinfo{pages}{053010} (\bibinfo{year}{2011}).

\bibitem[{\citenamefont{Lunde et~al.}(2013)\citenamefont{Lunde, Lopez-Monis,
  Vasiliadou, Bonilla, and Platero}}]{Platero_2}
\bibinfo{author}{\bibfnamefont{A.~M.} \bibnamefont{Lunde}},
  \bibinfo{author}{\bibfnamefont{C.}~\bibnamefont{Lopez-Monis}},
  \bibinfo{author}{\bibfnamefont{I.~A.} \bibnamefont{Vasiliadou}},
  \bibinfo{author}{\bibfnamefont{L.~L.} \bibnamefont{Bonilla}},
  \bibnamefont{and} \bibinfo{author}{\bibfnamefont{G.}~\bibnamefont{Platero}},
  \bibinfo{journal}{Phys. Rev. B} \textbf{\bibinfo{volume}{88}},
  \bibinfo{pages}{035317} (\bibinfo{year}{2013}).

\bibitem[{\citenamefont{Danon et~al.}(2009)\citenamefont{Danon, Vink, Koppens,
  Nowack, Vandersypen, and Nazarov}}]{Danon_PRL}
\bibinfo{author}{\bibfnamefont{J.}~\bibnamefont{Danon}},
  \bibinfo{author}{\bibfnamefont{I.~T.} \bibnamefont{Vink}},
  \bibinfo{author}{\bibfnamefont{F.~H.~L.} \bibnamefont{Koppens}},
  \bibinfo{author}{\bibfnamefont{K.~C.} \bibnamefont{Nowack}},
  \bibinfo{author}{\bibfnamefont{L.~M.~K.} \bibnamefont{Vandersypen}},
  \bibnamefont{and} \bibinfo{author}{\bibfnamefont{Y.~V.}
  \bibnamefont{Nazarov}}, \bibinfo{journal}{Phys. Rev. Lett.}
  \textbf{\bibinfo{volume}{103}}, \bibinfo{pages}{046601}
  (\bibinfo{year}{2009}).

\bibitem[{\citenamefont{Beenakker}(1991)}]{Beenakker}
\bibinfo{author}{\bibfnamefont{C.~W.~J.} \bibnamefont{Beenakker}},
  \bibinfo{journal}{Phys. Rev. B} \textbf{\bibinfo{volume}{44}},
  \bibinfo{pages}{1646} (\bibinfo{year}{1991}).

\bibitem[{\citenamefont{Muralidharan et~al.}(2006)\citenamefont{Muralidharan,
  Ghosh, and Datta}}]{Basky_Beenakker}
\bibinfo{author}{\bibfnamefont{B.}~\bibnamefont{Muralidharan}},
  \bibinfo{author}{\bibfnamefont{A.~W.} \bibnamefont{Ghosh}}, \bibnamefont{and}
  \bibinfo{author}{\bibfnamefont{S.}~\bibnamefont{Datta}},
  \bibinfo{journal}{Phys. Rev. B} \textbf{\bibinfo{volume}{73}},
  \bibinfo{pages}{155410} (\bibinfo{year}{2006}).

\bibitem[{\citenamefont{Timm}(2008)}]{Timm}
\bibinfo{author}{\bibfnamefont{C.}~\bibnamefont{Timm}}, \bibinfo{journal}{Phys.
  Rev. B} \textbf{\bibinfo{volume}{77}}, \bibinfo{pages}{195416}
  (\bibinfo{year}{2008}).

\bibitem[{\citenamefont{Fischer et~al.}(2008)\citenamefont{Fischer, Coish,
  Bulaev, and Loss}}]{Fischer}
\bibinfo{author}{\bibfnamefont{J.}~\bibnamefont{Fischer}},
  \bibinfo{author}{\bibfnamefont{W.~A.} \bibnamefont{Coish}},
  \bibinfo{author}{\bibfnamefont{D.~V.} \bibnamefont{Bulaev}},
  \bibnamefont{and} \bibinfo{author}{\bibfnamefont{D.}~\bibnamefont{Loss}},
  \bibinfo{journal}{Phys. Rev. B} \textbf{\bibinfo{volume}{78}},
  \bibinfo{pages}{155329} (\bibinfo{year}{2008}).

\bibitem[{\citenamefont{Braun et~al.}(2004)\citenamefont{Braun, K{\"o}nig, and
  Martinek}}]{Koenig}
\bibinfo{author}{\bibfnamefont{M.}~\bibnamefont{Braun}},
  \bibinfo{author}{\bibfnamefont{J.}~\bibnamefont{K{\"o}nig}},
  \bibnamefont{and} \bibinfo{author}{\bibfnamefont{J.}~\bibnamefont{Martinek}},
  \bibinfo{journal}{Phys. Rev. B} \textbf{\bibinfo{volume}{70}},
  \bibinfo{pages}{195345} (\bibinfo{year}{2004}).

\bibitem[{\citenamefont{Muralidharan and Grifoni}(2013)}]{Basky_Milena}
\bibinfo{author}{\bibfnamefont{B.}~\bibnamefont{Muralidharan}}
  \bibnamefont{and} \bibinfo{author}{\bibfnamefont{M.}~\bibnamefont{Grifoni}},
  \bibinfo{journal}{Phys. Rev. B} \textbf{\bibinfo{volume}{88}},
  \bibinfo{pages}{045402} (\bibinfo{year}{2013}).

\bibitem[{\citenamefont{Hu and Wang}(2013)}]{Osc1}
\bibinfo{author}{\bibfnamefont{B.}~\bibnamefont{Hu}} \bibnamefont{and}
  \bibinfo{author}{\bibfnamefont{X.~R.} \bibnamefont{Wang}},
  \bibinfo{journal}{Phys. Rev. B} \textbf{\bibinfo{volume}{87}},
  \bibinfo{pages}{035311} (\bibinfo{year}{2013}).

\bibitem[{\citenamefont{Rudner and Levitov}(2013)}]{Osc2}
\bibinfo{author}{\bibfnamefont{M.~S.} \bibnamefont{Rudner}} \bibnamefont{and}
  \bibinfo{author}{\bibfnamefont{L.~S.} \bibnamefont{Levitov}},
  \bibinfo{journal}{Phys. Rev. Lett.} \textbf{\bibinfo{volume}{110}},
  \bibinfo{pages}{086601} (\bibinfo{year}{2013}).

\end{thebibliography}

\end{document}